\documentclass{article}
\usepackage{neurips_2025}

\usepackage[utf8]{inputenc} 
\usepackage[T1]{fontenc}    
\usepackage{hyperref}       
\usepackage{url}            
\usepackage{booktabs}       
\usepackage{amsfonts}       
\usepackage{amsmath}        
\usepackage{nicefrac}       
\usepackage{microtype}      
\usepackage{xcolor}         
\usepackage{graphicx}
\usepackage{epstopdf}
\usepackage{algorithm}      
\usepackage{algpseudocode} 
\usepackage{makecell} 

\title{UniTTS: An end-to-end TTS system without decoupling of acoustic and semantic information}

\begin{document}
\author{
 \textbf{\small Rui Wang\textsuperscript{1,3}}\thanks{These authors contributed equally to this work.},
 \textbf{\small Qianguo Sun\textsuperscript{1}\footnotemark[1]},
 \textbf{\small Tianrong Chen\textsuperscript{2},}
 \textbf{\small Zhiyun Zeng\textsuperscript{2},}
 \textbf{\small Junlong Wu\textsuperscript{2},}
 \textbf{\small Jiaxing Zhang\textsuperscript{3}\thanks{Corresponding author.}}
\\
\small{\textsuperscript{1}}\small{International Digital Economy Academy, IDEA\&Emdoor Collaborative Laboratory} \\
\small{\textsuperscript{2}}\small{Emdoor Information Co., Ltd} \\
\small{\textsuperscript{3}}\small{Shenzhen Yijiayiban Information Technology Co., Ltd}
\\
 \texttt{\{wangrui, sunqianguo\}@idea.edu.cn} \\
 \texttt{\{wangrui, zhangjiaxing\}@aiipal.com.cn} \\
 \texttt{\{tianrong.chen, zhiyun.zeng, junlong.wu\}@emdoor.com}\\
}

\maketitle

\begin{abstract}
  The emergence of multi-codebook neutral audio codecs such as Residual Vector Quantization (RVQ) and Group Vector Quantization (GVQ) has significantly advanced Large-Language-Model (LLM) based Text-to-Speech (TTS) systems. These codecs are crucial in separating semantic and acoustic information while efficiently harnessing semantic priors. However, since semantic and acoustic information cannot be fully aligned, a significant drawback of these methods when applied to LLM-based TTS is that large language models may have limited access to comprehensive audio information. To address this limitation, we propose DistilCodec and UniTTS, which collectively offer the following advantages: 1) This method can distill a multi-codebook audio codec into a single-codebook audio codec with 32,768 codes while achieving a near 100\% utilization. 2) As DistilCodec does not employ a semantic alignment scheme, a large amount of high-quality unlabeled audio (such as audiobooks with sound effects, songs, etc.) can be incorporated during training, further expanding data diversity and broadening its applicability. 3) Leveraging the comprehensive audio information modeling of DistilCodec, we integrated three key tasks into UniTTS's pre-training framework: audio modality autoregression, text modality autoregression, and speech-text cross-modal autoregression. This allows UniTTS to accept interleaved text and speech/audio prompts while substantially preserving LLM's text capabilities. 4) UniTTS employs a three-stage training process: Pre-Training, Supervised Fine-Tuning (SFT), and Alignment. Source code and model checkpoints are publicly available at \href{https://github.com/IDEA-Emdoor-Lab/UniTTS}{https://github.com/IDEA-Emdoor-Lab/UniTTS} and \href{https://github.com/IDEA-Emdoor-Lab/DistilCodec}{https://github.com/IDEA-Emdoor-Lab/DistilCodec}.
\end{abstract}

\section{Introduction}
Recent years have witnessed remarkable advancements in Large Language Models (LLMs)\cite{radford2019language,kaplan2020scaling,grattafiori2024llama,yang2024qwen2}. Their exceptional capability in discrete token modeling has garnered significant attention in the speech processing community. With the development of multimodal discretization techniques such as Vector Quantization (VQ)\cite{van2017neural}, Finite Scalar Quantization (FSQ)\cite{mentzer2023finite}, and Grouped-Residual-Factorized Vector Quantization (GRFVQ, including Grouped-VQ, Residual-VQ, and Factorized-VQ)\cite{zeghidour2021soundstream,yang2023hifi,yu2021vector}, the performance of Neural Audio Codecs (NAC) has been substantially improved. Consequently, an increasing number of Text-To-Speech (TTS) systems are adopting LLM-based approaches \cite{du2024cosyvoice2,du2024cosyvoice,wang2025spark,ye2025llasa,liao2024fish,deng2025indextts,chen2024f5,anastassiou2024seed}, which demonstrate superior performance in both speech naturalness and emotional expressiveness.

The performance of LLM-based text-to-speech (TTS) systems is heavily influenced by the discrete audio tokens from by NACs. Recent research has shown that integrating semantic distillation\cite{zhang2023speechtokenizer,defossez2024moshi,ye2025codec,ye2025llasa,wang2025spark} into NAC can significantly enhance TTS system’s performance by leveraging rich semantic information from audio encoders \cite{baevski2020wav2vec,radford2023robust}. While semantic distillation has significantly improved semantic representation capabilities for LLM-based TTS systems, not all speech can be decomposed into discrete semantic and acoustic features. This limitation is particularly evident in prosodically salient non-linguistic vocalizations (e.g., laughter or crying) and high-fidelity universal audio containing acoustically complex backgrounds or layered sound effects. While some NAC implementations employ GRFVQ-based multi-codebook architectures to enhance performance, such approaches result in substantially increased bitrates in the discretized speech sequences. These elongated sequences introduce substantial modeling challenges for language models (LLMs) in capturing speech relationships, thereby motivating the development of efficient low-bitrate NAC variants\cite{li2024single,xin2024bigcodec,parker2024scaling,wang2025spark}. These observations lead to three critical research questions: (i) How to develop low-bitrate, high-performance NACs for universal audio without relying on semantic or additional prior information? (ii) How to optimally utilize universal audio data in NAC and LLM-based TTS training? (iii) Whether LLMs coupled with universal audio NACs can effectively achieve text-audio alignment?

\begin{figure}
\centering
\includegraphics[scale=0.5]{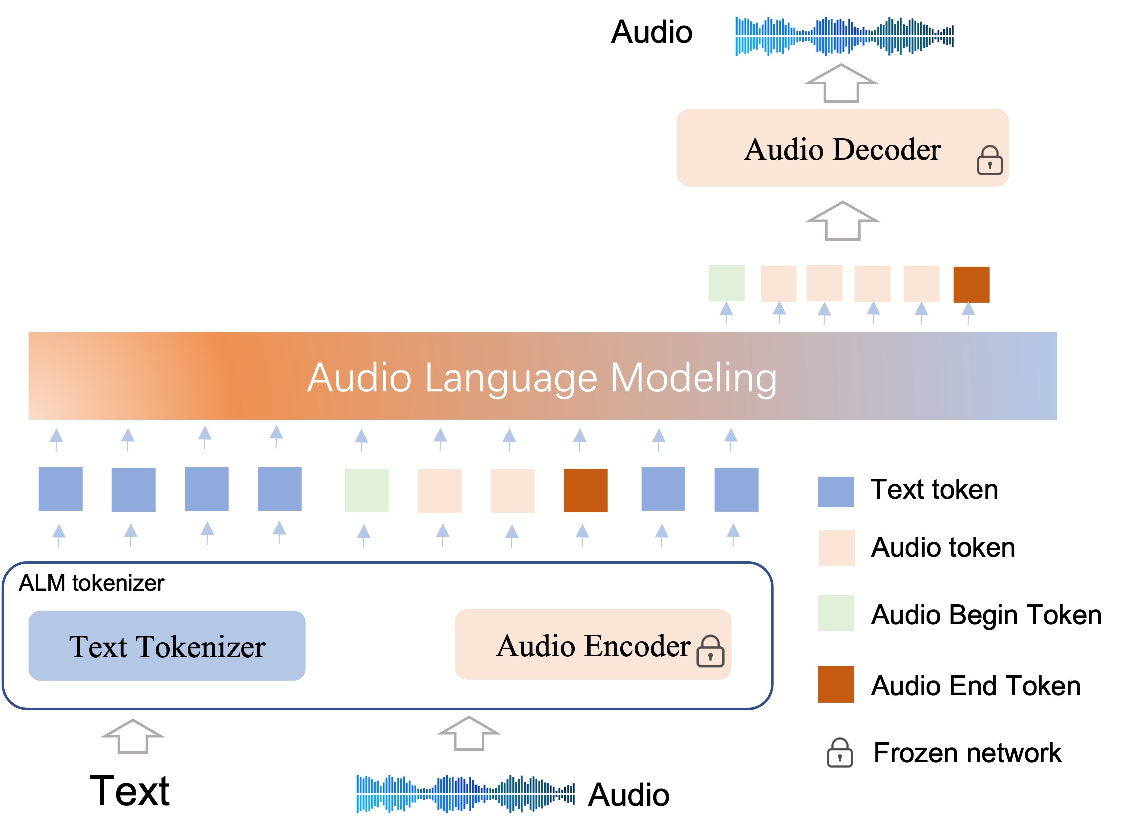}
\caption{The UniTTS architecture consists of an ALM tokenizer and an ALM backbone network, supporting both text and audio inputs and outputs. Within the architecture, DistilCodec is responsible for audio signal transformation: its encode module discretizes audio into latent representations, while the decode module reconstructs the waveform for acoustic output.}
\label{figure_1_framwork}
\end{figure}

Based on the motivation outlined previously, we introduce DistilCodec and UniTTS. DistilCodec is a single-codebook audio codec, which has 32768 codes, and the utilization of the codebook achieves nearly 100\%. UniTTS leverages DistilCodec for audio discretization, while its backbone network adopts Qwen2.5-7B\cite{yang2024qwen2} to model relationships between audio tokens. The architecture of UniTTS is illustrated in Figure \ref{figure_1_framwork}.
Our main contributions are summarized as follows:
\begin{itemize}
  \item \textbf{DistilCodec}: We propose a training methodology that enables the distillation of multi-codebook NAC into single-codebook NAC. Through this approach, we have developed DistilCodec - a single-codebook NAC containing 32,768 codes that achieves 100\% utilization with balanced code distribution. Notably, DistilCodec employs universal audio data for training rather than being restricted to speech-specific datasets.
  \item \textbf{UniTTS}: We present UniTTS, a novel TTS system trained on QWen2.5-7B and DistilCodec. Leveraging DistilCodec's comprehensive audio modeling capabilities, UniTTS achieves end-to-end speech synthesis with full-spectrum audio input/output. The system demonstrates enhanced naturalness in emotional expressiveness compared to conventional TTS systems, particularly in capturing subtle prosodic variations and affective nuances during audio generation. 
  \item \textbf{Novel Audio Language Model Paradigm}: We establish a dual-phase Audio Language Model (ALM) training framework, which comprises (i) Audio Perceptual Modeling (DistilCodec) focusing purely on acoustic discretization, and (ii) Audio Cognitive Modeling (UniTTS) implemented via pretraining (incorporating universal audio autoregressive tasks), supervised fine-tuning (evaluating text-audio interleaved prompts' impact), and alignment (employing direct preference optimization for speech refinement) - enabled by UniTTS's complete end-to-end integration within the LLM.
\end{itemize}

\section{Related Work}
\subsection{NAC (Neural Audio Codec)}
Recently, a major research focus in NAC lies in effectively integrating prior information, such as semantic features. SpeechTokenizer\cite{zhang2023speechtokenizer}, building upon RVQ\cite{zeghidour2021soundstream}, pioneered semantic alignment between HuBERT\cite{hsu2021hubert} representations and the first RVQ layer through model distillation. Following this paradigm, subsequent works including Mimi\cite{defossez2024moshi}, X-codec\cite{ye2025codec}, X-codec2\cite{ye2025llasa}, and BiCodec\cite{wang2025spark}. Another critical research direction involves achieving low bitrate while maintaining high performance, with representative works encompassing Single-Codec\cite{li2024single}, StableCodec\cite{parker2024scaling}, BiCodec\cite{wang2025spark}, X-codec / X-codec2\cite{xin2024bigcodec,ye2025llasa}, and WavTokenizer\cite{ji2024wavtokenizer}. While WavTokenizer and our DistilCodec both employ low-bitrate single-codebook NAC with full audio modeling, WavTokenizer uses a small 4k codebook. Prior works\cite{ma2025unitok,zhu2024scaling} suggest that codebook size significantly affects NAC's performance. In the field of NAC, StableCodec with a codebook size of 15,625 and X-codec2 featuring a codebook size of 65,536 have both demonstrated the significant performance improvements achieved by large codebooks in speech discretization tasks.

\subsection{LLM-based TTS}
VALL-E\cite{wang2023neural} pioneered the first LLM-based text-to-speech (TTS) framework, utilizing Encodec\cite{defossez2022high} as its audio tokenizer and adopting a multi-stage decoding strategy that combines non-autoregressive (NAR) and autoregressive (AR) generation. However, this approach results in incomplete audio information being processed by the LLM and introduces significant decoding complexity. In contrast, MELL-E\cite{meng2024autoregressive} and KALL-E\cite{zhu2024autoregressive} directly employ continuous acoustic features as input, demonstrating the viability of non-semantically aligned TTS frameworks. Nonetheless, these methods face scalability limitations in large-scale training scenarios.

To address the high computational overhead of multi-stage decoding, recent work has shifted toward single-stage paradigms. For example, Llasa\cite{ye2025llasa} exploits the unified semantic-acoustic codebook of X-codec2, empirically validating the scaling laws for both data and model size in TTS tasks. Similarly, Spark-TTS\cite{wang2025spark} integrates speaker characteristics and speech tokens for synthesis, while CosyVoice1.0\cite{du2024cosyvoice} and CosyVoice2.0\cite{du2024cosyvoice2} decompose speech into speaker embeddings and semantic tokens, subsequently generating waveforms via flow matching.

\section{Methods}
\subsection{Overview}
UniTTS distinguishes itself from existing LLM-based TTS systems by introducing DistilCodec, an audio tokenizer capable of holistically modeling universal audio signals. This design eliminates information loss inherent in semantic alignment while substantially reducing decoding complexity through a streamlined single-stage architecture.

The overall architecture of UniTTS consists of three key components: (1) The Encoder and Quantizer of DistilCodec serving as the Audio Tokenizer; (2) A Large Language Model (LLM) utilizing QWen2.5-7B\cite{yang2024qwen2}; (3) The Decoder of DistilCodec functioning as the speech synthesizer. The training process of UniTTS is divided into two stages(detailed training schema illustrated in Appendix \ref{app:unitts}):
\begin{itemize}
  \item \textbf{Audio Perceptual Modeling}: We have developed a novel distillation approach termed DMS (\textbf{D}istilling \textbf{M}ulti-Codebook NAC to \textbf{S}ingle-Codebook NAC) by enabling the Student NAC to inherit encoder and decoder parameters from the Teacher NAC. Based on DMS, we trained DistilCodec using universal audio datasets as training data, achieving a single codebook with a codebook size of 32,768 while maintaining codebook utilization approaching 100\%. Simultaneously, the DMS algorithm enables the dimension of the distilled Student NAC Codebook to be scaled beyond 2048. Leveraging this capability, we configured the codebook dimension to 3584, aligning with the word embedding dimension of QWen2.5-7B (3584), so we subsequently leveraged DistilCodec's codebook to initialize the audio embedding layer in UniTTS.
  \item \textbf{Audio Cognitive Modeling}: The Codebook of DistilCodec is concatenated with the Word Embedding of QWen2.5, resulting in a vocabulary size of approximately 180,000 for UniTTS. Given UniTTS's capability to process complete audio information, its training paradigm mirrors that of Large Language Models (LLMs), divided into three stages: Pretrain, Supervised Fine-Tuning (SFT), and Alignment. During the pretraining phase, we incorporated an additional audio autoregressive task leveraging the universal audio modeling capability of DistilCodec. To preserve the textual capabilities of QWen2.5-7B, we simultaneously retained a portion of text data for training continuation. The SFT phase involved empirical validation of various audio-text interleaved prompt formats to optimize performance. Finally, the Alignment stage employed Direct Preference Optimization (DPO) to further enhance the quality of speech generation outputs.
\end{itemize}

\subsection{Audio Perception Modeling: DistilCodec}
The foundational network architecture of DistilCodec adopts an Encoder-VQ-Decoder framework similar to that proposed in Soundstream\cite{zeghidour2021soundstream}. The encoder employs a ConvNeXt-V2\cite{woo2023convnextv2codesigningscaling} structure, while the vector quantization module implements the GRFVQ scheme \cite{yang2023hifi,yu2021vector,zeghidour2021soundstream}. The decoder employs a ConvTranspose1d based  architectural configuration similar to HiFiGAN\cite{kong2020hifi}. Detailed network specifications and layer configurations are provided in Appendix \ref{app:distilcodec_structure}
The training methodology of DistilCodec follows a similar approach to HiFiGAN \cite{kong2020hifi, qi2018losssensitivegenerativeadversarialnetworks}, incorporating three types of discriminators: Multi-Period Discriminator (MPD), Multi-Scale Discriminator (MSD), and Multi-STFT Discriminator (MSFTFD). The detailed parameter configurations for these discriminators can be found in Appendix 
 \ref{app:discriminators}, and the comprehensive training diagram of DisitilCodec is provided in Appendix \ref{app:DTF}. The training objectives are formulated as follows:
\begin{equation}
\centering
L_{\text{total}} = \lambda_{\text{mel}} L_{\text{mel}}(G) + L_{\text{adv}}(G, D) + \lambda_{\text{fm}} L_{\text{FM}}(G, D)
\label{eq_gan_loss}
\end{equation}
The Mel-spectrogram loss function is denoted as $\lambda_{\text{mel}} L_{\text{mel}}(G)$, the loss functions for the three discriminators are represented by $L_{\text{adv}}(G, D)$, while their corresponding feature map loss functions are designated as $\lambda_{\text{fm}} L_{\text{FM}}(G, D)$. The operational workflow of the DisilCodec-LSGAN-Training (DLF) is presented in Appendix\ref{app:DTF}.
The training process of DistilCodec consists of two distinct phases: teacher-training and student-training. We abbreviate the NAC in each phase as:
\begin{equation}
\centering
Codec_{sx} = M(N_r, N_g, N_c, N_{dim}, E_{param}^{from}, G_{param}^{from}, VQ_{param}^{from})
\label{eq_codec_model}
\end{equation}
$N_r$ denotes the number of NAC residual layers, $N_g$ represents the quantity of NAC Groups, and $N_{\text{dim}}$ indicates the dimension of the NAC Codebook. The parameters $E_{param}^{from}$ are initialized from the encoder of a specific codec, $G_{param}^{from}$ initialized from the Generator/Decoder of a codec, and $VQ_{param}^{from}$ correspond to the Vector Quantization (VQ) parameters from a codec. The detailed NAC architecture settings used in our training are presented in Table \ref{codec_setting}. 
Under the framework of the DMS (\textbf{D}istilling \textbf{M}ulti-Codebook NAC to \textbf{S}ingle-Codebook NAC), we first trained the $\textbf{Teacher}_{\text{Codec}}$ using DLF, then trained the $\textbf{Student}_{\text{Codec}}$ (namely DisilCodec) through parameter inheritance from both Encoder and Decoder of $\textbf{Teacher}_{\text{Codec}}$. The pseudo-code of DMS is presented in Algorithm \ref{alg_dms}.
\begin{table}[h]
\caption{Settings of Two Stage NAC}
\label{tab:codec_comparison}
\centering
\begin{tabular}{l|cccc}
\toprule
\textbf{Codec} & \textbf{N-Residual} & \textbf{N-Group} & \textbf{N-Codes/Codebook} & \textbf{Dimension} \\ 
\midrule
Teacher-Codec & 8 & 4 & 1024 & 512 \\
Student-Codec & 1 & 1 & 32768 & 3584 \\
\bottomrule
\end{tabular}
\label{codec_setting}
\end{table}

\begin{algorithm}
\label{alg_dms}
\caption{DMS: Distilling Multi-Codebook NAC to Single-Codebook NAC via parameter inheritance)}
\begin{algorithmic}[1]
\State \textbf{Step 1:} Initializing $\text{Teacher}_{\text{codec}}$:
\Statex \hspace*{\algorithmicindent} $\text{Teacher}_{\text{codec}} = \text{Codec}_{s1}(8, 4, 1024, 512, E^{\text{scratch}}_{\text{param}}, G^{\text{scratch}}_{\text{param}}, VQ^{\text{scratch}}_{\text{param}})$
\State \textbf{Step 2:} $\text{Teacher}_{\text{codec}}$ training with DLF
\State \textbf{Step 3:} Initializing $\text{Student}_{\text{codec}}$:
\Statex \hspace*{\algorithmicindent} $\text{Student}_{\text{codec}} = \text{Codec}_{s2}(1, 1, 35768, 3584, E^{\text{teacher}}_{\text{param}}, G^{\text{teacher}}_{\text{param}}, VQ^{\text{scratch}}_{\text{param}})$
\State \textbf{Step 4:} $\text{Student}_{\text{codec}}$ training with DLF
\State \textbf{Output:} $\text{DistilCodec} = \text{Student}_{\text{codec}}$
\end{algorithmic}
\end{algorithm}

\subsection{Audio Cognitive Modeling}
\subsubsection{Pretrain}
In the pre-training stage, integrating the audio modality into the language model can be equivalently formulated as modeling $\text{P}(A, T)$, where $A$ denotes audio and $T$ denotes text. Consequently, the joint modeling of audio and text can be decomposed using Bayes' theorem, as shown in the following equation:
\begin{equation}
p(A \cdot T) = p(A|T) \cdot p(T) = p(T|A) \cdot p(A)
\end{equation}
From this decomposition, three distinct pretrain tasks of UniTTS can be derived:
\begin{itemize}
  \item $\textbf{p(A)}$: Audio Modality Auto-regression with universal audio data
  \item $\textbf{p(T)}$: Text Modality Auto-regression
  \item $\textbf{p(A|T)} \& \textbf{p(T|A)}$: Text-Audio Modality Conditional Modeling
\end{itemize}

Consequently, our audio modality pre-training extends the text modality by incorporating audio modality auto-regression and text audio alignment tasks. Additionally, as demonstrated in Appendix \ref{PA_enhance_audio}, our experiments reveal that audio modeling presents greater spatial complexity and implementation challenges compared to text modeling. The scarcity of high-quality text-audio paired data further necessitates the integration of a universal audio autoregressive task, which proves beneficial for enhancing final model performance. Within this pre-training phase, we designed a multi-stage training approach.

\begin{itemize}
  \item Stage 1: The model was trained on text data, universal audio data, and a limited amount of paired text-audio data to learn relationships for audio modeling. However, when audio training data was introduced into a model initialized from a pre-trained text model, modal competition occurred, leading to observable degradation in the model’s text generation capability. This outcome directly motivated the subsequent training in Stage 2.
  \item Stage 2: We augmented the training data by incorporating text-based instruction datasets alongside the existing universal audio and text-audio pair datasets, further enhancing the model's text generation capabilities (see Appendix \ref{Text_Capability_Testing_of_UniTTS} for details). Additionally, to accommodate longer contextual sequences, we expanded the model's context window size from 8,192 to 16,384 during this stage.
  
\end{itemize}

\subsubsection{SFT}
The quality of instruction tuning data significantly impacts the final model performance. However, existing open-source text-speech aligned datasets exhibit two primary limitations: 1) The corresponding text is often derived from ASR model outputs, inherently containing noise. 2) Many audio samples are extracted from sources such as podcasts and audiobooks, frequently featuring excessively long silent segments, which are suboptimal for TTS model training. To address these issues and enhance semantic alignment quality, we implemented a data filtering algorithm based on a composite quality score. This score is derived from metrics assessing both text accuracy and audio quality. Text accuracy is evaluated by generating reference text using the Paraformer\cite{gao2022paraformer} and Whisper\cite{radford2023robust} models and computing the Character Error Rate (CER). Audio quality is assessed using the DNSMOS P.835 OVRL metric\cite{reddy2022dnsmos}. Based on these combined indicators, we compute a novel composite score for each data sample
\begin{equation}
\text{quality}(i) = \text{dnsmos}(i) - \text{cer}(i)
\end{equation}
Where i denotes the index of the sample, quality(i) represents the DNSMOS score for sample i, and cer(i) represents the CER score for sample i. Samples are subsequently ranked in descending order based on their quality scores, and a predetermined number of the highest-ranked samples are selected for inclusion in the training set. The pseudocode is presented in Appendix \ref{app:data_filtering}.

Our experiments demonstrated that incorporating text instruction data and the long-cot instruction dataset not only enhanced the model's text understanding capability but also led to further improvements in audio generation quality. The templates for text-to-speech (TTS) conversion, text dialogue, and long-CoT instructions are provided in Appendix \ref{app:prompt_template}.

\subsubsection{Alignment}
Following instruction fine-tuning, the model often exhibits issues such as prosodic lengthening and repetition. These phenomena are analogous to the text repetition problems observed in Large Language Models (LLMs) that have undergone insufficient pre-training. We therefore hypothesize that these issues stem from insufficient pre-training data in the speech modality. As evidenced by the pre-training loss curve in Fig. \ref{figure_8_training_loss}, the audio generation loss persists at elevated levels while maintaining a consistent downward trajectory.

To further enhance the stability of the model's audio generation, we employed Direct Preference Optimization (DPO)\cite{rafailov2023direct}. However, when applied to ultra-long sequence tasks such as UniTTS, the vanilla DPO algorithm is susceptible to mode collapse. Consequently, to address these challenges, we employ linear preference optimization (LPO) as an alternative to DPO. The training objective of LPO is:
\begin{equation}
\label{lpo_eq}
L_{\text{lpo}} = \gamma \cdot (x_1^{\text{ste}} + x_2^{\text{ste}}) + \lambda \max(0, -\log x_1^{\text{ste}})
\end{equation}
In formula \ref{lpo_eq}, $x_1^{\text{ste}}$, $x_2^{\text{ste}}$, $\gamma$, $x_1$, $x_2$ are detailed in Appendix \ref{app:lpo}.

\subsection{Evaluation}

Since DistilCodec is trained on universal audio, to verify whether the model can generate more natural and emotionally expressive outputs based on the understanding of semantics, we referenced Llasa\cite{ye2025llasa} to construct a diversity evaluation dataset. This includes six main emotions, rare characters, tongue twisters, interjections, audiobooks, and special casual conversation scenarios, such as stuttering. To comprehensively assess vocal timbre replication capabilities across demographic variations, we specifically curated voice samples covering four critical age-gender categories: children, male adults, female adults, and elderly speakers.

\section{Experiments}
This section first evaluates the performance of the proposed DistilCodec, followed by an assessment of UniTTS.

\subsection{Experimental Setup for DistilCodec}
For the training of DistilCodec, we employed a diverse corpus of 100,000 hours of audio data spanning multiple domains, including Chinese audiobooks, Chinese speech, English audiobooks, English speech, musical compositions, and sound effects. Detailed data distributions are presented in Table \ref{tab:codec_data_size}.

The training of DisitilCodec utilized the computational resources of 5×8 A100 GPUs. Specifically, the Teacher Codec was trained for 5 epochs, while the Student Codec was trained for 3 epochs in the second training stage. Both the Teacher Codec and the Student Codec were trained using AdamW as the optimizer. Parameters for AdamW are detailed in Appendix \ref{app:DTF}. 

\subsection{DistilCodec Evaluation}
First, we evaluate the codebook utilization rate of DistilCodec, measured using Codebook Perplexity (PPL). The formula for calculating PPL is as follows:
\begin{equation}
\centering
\label{cb_ppl}
\text{Perplexity} = e^{-\sum_{i=1}^{K} p_i \log p_i}  
\end{equation}
Let $p_i = \frac{N_i}{N_{total}}$, denote the probability of selecting the i-th codebook entry during quantization, where K is the size of the codebook.

We employed the LibriSpeech-Clean dataset and a self-constructed Universal Audio dataset to evaluate the perplexity and codebook utilization rate of DistilCodec, with the corresponding experimental results presented in Table \ref{tab:ppl_cmp}. From the Table \ref{tab:ppl_cmp}, it can be observed that DistilCodec achieves near-optimal codebook utilization (approaching 100\%) across both speech and universal audio datasets. However, due to the increased complexity of universal audio, the perplexity (PPL) of the codebook demonstrates higher values for universal audio compared to speech. Additionally, we conducted a comprehensive comparative analysis of DistilCodec's speech reconstruction capabilities using the LibriSpeech-Clean-Test benchmark, and the results are shown in Table \ref{tab:comparison}.
\begin{table}[h]
\caption{Comparison of model performance metrics}
\label{tab:ppl_cmp}
\centering
\begin{tabular}{l|cc}
\toprule
\textbf{Dataset} & \textbf{Codebook Usage(\%)↑} & \textbf{Codebook PPL↑} \\ 
\midrule
LibriSpeech-Clean-Test & 98.2 & 21660.5 \\
Universal-Audio-Test  & 99.9 & 26999.0 \\ 
\bottomrule
\end{tabular}
\end{table}

\begin{table}[h]
\caption{Comparison of various models based on codebook size, token rate, bandwidth, and quality metrics}
\label{tab:comparison}
\centering
\begin{tabular}{l|cccc|ccc}
\toprule
\makecell{\textbf{Model}}
& \makecell{\textbf{Codebook}\\\textbf{Size}} & 
\makecell{\textbf{Nq}} & 
\makecell{\textbf{Token Rate}\\\textbf{(TPS)}} & 
\makecell{\textbf{Bandwidth}\\\textbf{(bps)}} & 
\makecell{\textbf{STOI}\\\textbf{↑}} & 
\makecell{\textbf{PESQ}\\\textbf{↑}} & 
\makecell{\textbf{UTMOS}\\\textbf{↑}} \\ 
\midrule
Encodec & 1024 & 8 & 600 & 6000 & 0.94 & 2.75 & 3.07 \\
DAC & 1024 & 12 & 600 & 6000 & 0.95 & 4.01 & 4.00 \\
Encodec & 1024 & 2 & 150 & 1500 & 0.84 & 1.56 & 1.58 \\
Mimi & 2048 & 8 & 100 & 1100 & 0.91 & 2.25 & 3.56 \\
BigCodec & 8192 & 1 & 80 & 1040 & 0.94 & 2.68 & 4.11 \\
DAC & 1024 & 2 & 100 & 1000 & 0.73 & 1.14 & 1.29 \\
SpeechTokenizer & 1024 & 2 & 100 & 1000 & 0.77 & 1.25 & 2.28 \\
X-codec & 1024 & 2 & 100 & 1000 & 0.86 & 2.33 & 4.21 \\
WavTokenizer & 4096 & 1 & 75 & 900 & 0.89 & 2.14 & 3.94 \\
X-codec2 & 65536 & 1 & 50 & 800 & 0.92 & 2.43 & 4.13 \\
StableCodec & 15625 & 2 & 50 & 697 & 0.91 & 2.24 & 4.23 \\
Single-Codec & 8192 & 1 & 23.4 & 304 & 0.86 & 1.88 & 3.72 \\
BiCodec & 8192 & 1 & 50 & 650 & 0.92 & 2.51 & 4.18 \\ 
\midrule
\textbf{DistilCodec} & 32768 & 1 & 93 & 1300 & 0.93 & 2.02 & 3.75 \\ 
\bottomrule
\end{tabular}
\end{table}

Since DistilCodec was trained on universal audio, we first employed UTMOS\cite{saeki2022utmos} for automatic quality assessment. However, the universal audio test set received an unreliable low score (1.89), indicating UTMOS's inadequacy for universal audio evaluation. We therefore conducted a Mean Opinion Score (MOS) evaluation, which consists of:

\textbf{Evaluation Dataset}: We selected 98 universal audio clips, comprising Chinese and English audiobooks, streaming media audio content, and sound effects.

\textbf{Evaluation Protocol}:
\begin{itemize}
  \item Speech Clarity: Subjective rating (0-5 scale) of vocal articulation quality.
  \item Background Clarity: Subjective rating (0-5 scale) of environmental sound.
\end{itemize}

Table \ref{tab:mos_score} presents the MOS results for universal audio under this evaluation framework. Evaluation results demonstrate that DistilCodec achieves superior scores in both speech clarity and background clarity, indicating its capability for universal audio reconstruction.
\begin{table}[h]
\caption{Comparison of MOS Scores}
\label{tab:mos_score}
\centering
\begin{tabular}{l|cc}
\toprule
\textbf{Assessment Items} & \textbf{Synthetic Speech} & \textbf{GT} \\ \midrule
Speech Clarity & 4.689 & 4.945 \\
Background Audio Clarity &4.768 & 4.927 \\
Average Score & 4.728 & 4.4.936 \\
\bottomrule
\end{tabular}
\end{table}

\subsection{TTS experiments}
\subsubsection{EXPERIMENTAL DETAILS}
\textbf{Pretraining}: Our pretraining corpus comprises three data modalities: (1) universal audio data, (2) text data, and (3) text-audio aligned data. During the pre-training phase, our model was trained on a total of 322B tokens, incorporating partial data from Libriheavy\cite{kang2024libriheavy}, WenetSpeech4TTS\cite{ma2024wenetspeech4tts}, Emilia\cite{he2024emilia}, as well as our proprietary datasets. The text data consists of our self-collected datasets, Infinity-Instruct, and SkyPile-150B. The complete data distribution is detailed in Appendix \ref{section:pertaining_data}.

The pretraining process consists of two stages: the first stage employs a cosine-annealed learning rate schedule decaying from 1e-4 to 2e-5 with 10\% warmup proportion, employing 8,192 windows and a batch size of 256 for training efficiency; the second stage continues with a finer learning rate decay from 2e-5 to 9e-6 while extending the window length to 16,384 and maintaining the same batch size. 

 \textbf{Supervised Fine-Tuning}: For fine-tuning, we employ a multi-task learning approach incorporating text instruction data, long-cot reasoning data, and TTS datasets to enhance the model's conversational understanding and audio generation capabilities, with the total audio data comprising approximately \textbf{948} hours (significantly lower than LLaSA's 250K hours and Spark-TTS's 100K hours) as detailed in Appendix \ref{section:sft_data_distrition}.

During the SFT phase, we adopt a cosine learning rate schedule that decays from 9e-6 to 5e-6, with a context window length of 8,192 and a batch size of 128.

\textbf{Alignment}: To further stabilize the model's performance, we employed LPO training. The distribution of the original alignment data used for LPO is provided in Appendix \ref{app:lpo}:

Next, following the preference pair generation method described in LPO, we used each sample’s prompt as input to generate three candidate responses. These candidates were then paired with the sample’s reference answer to form three preference pairs, which constitute the training data for LPO. The training parameters for LPO are listed in Table \ref{tab:dpo_param_values}.

\subsubsection{Text-to-Speech System Evaluation}
Our proposed UniTTS utilizes DistilCodec to extract comprehensive audio information, jointly modeling both acoustic features and semantic representations, unlike existing approaches that treat these aspects separately. We evaluate the audio generation performance under this unified framework, employing a three-stage training pipeline comprising pretraining, SFT, and alignment. We denote the model after SFT as UniTTS-SFT and the post-LPO-trained model as UniTTS-LPO.
For a rigorous evaluation, we compared UniTTS against state-of-the-art methods, including:CosyVoice2, Spark-TTS, LLaSA, F5-TTS\cite{liao2024fish}, Fish-Speech\cite{chen2024f5}, IndexTTS\cite{deng2025indextts}. 

The evaluation methodology is detailed in the Appendix \ref{app:tts_evaluation_criteria}, and the experimental results are presented in Table \ref{tab:tts_model_comparison}.

\begin{table}[h]
    \caption{Comparison of Mean Opinion Score between different TTS models.}
    \label{tab:tts_model_comparison}
    \centering
    \begin{tabular}{l|cccc}
        \toprule
        \textbf{Model} & \textbf{Fidelity} & \textbf{Stability} & \textbf{Naturalness} & \textbf{Emotional expressiveness} \\
        \midrule
        Cosyvoice2 & 4.80 & 5 & 4.89 & 4.11 \\
        SparkTTS & 4.89 & 5 & 4.89 & 4.26 \\
        Llasa & 4.74 & 4.91 & 4.91 & 4.11 \\
        F5-TTS & 4.94 & 5 & 4.89 & 3.97 \\
        Fish Speech & 4.89 & 5 & 4.83 & 4.29 \\
        IndexTTS & 4.69 & 4.83 & 4.89 & 4.31 \\
        UniTTS-SFT & 4.43 & 5 & 4.77 & 4.23 \\
        UniTTS-LPO & 4.80 & 4.97 & 4.94 & 4.60 \\
        \bottomrule
    \end{tabular}
\end{table}

The results presented in Table \ref{tab:tts_model_comparison} demonstrate that UniTTS-LPO achieves comprehensive improvements over UniTTS-SFT in emotional expressiveness, fidelity, and naturalness, thereby validating the effectiveness of the LPO training methodology. UniTTS-LPO's performance superiority stems from its DistilCodec-powered holistic modeling of prosodic-timbral-emotional features and diverse unsupervised training, enabling state-of-the-art emotion-aware speech synthesis.

\subsubsection{Ablation Analysis} \label{ablation_test}
To investigate the factors influencing the performance of UniTTS, we conducted comprehensive ablation experiments. We employed the CER (Character Error Rate) metric from seed-tts-eval as our evaluation criterion for the ablation study.
\begin{table}[h]
\caption{Comparison of CER for different ablation tests}
\label{tab:ablation_cer}
\centering
\begin{tabular}{l|c}
\toprule
\textbf{Ablation TestID} & \textbf{CER} \\ \midrule
AB-1 & 3.466\% \\ 
AB-2 & 5.5045\% \\ 
AB-3 & 3.582\% \\ 
AB-4 & 3.7025\% \\ 
AB-5 & 3.7395\% \\
\bottomrule
\end{tabular}
\end{table}

\textbf{Impact of Text Instructions on Model Performance:}
In Experiment AB-1, the TTS instruction dataset was augmented with text dialogue and long-CoT text instruction datasets, while AB-4 solely used the single-task TTS training dataset. A comparison between AB-1 and AB-4 demonstrates that incorporating text instructions improves audio generation quality.

\textbf{Influence of Instruction Templates on Model Performance:}
Compared to AB-1, AB-2’s prompt template only provided example audio without the corresponding reference text. As shown in Table \ref{tab:ablation_cer}, including both the example audio and its associated text in the prompt template yields a performance improvement of approximately 2.1\%. Furthermore, a comparison between AB-5 and AB-1 reveals that the ordering of the example audio and its corresponding text in the prompt template also has a measurable impact on performance.

\textbf{Compatibility Between TTS and ASR Tasks:}
In contrast to AB-1, AB-3 incorporated an ASR instruction dataset during training. During inference, we observed that some generated audio sequences contained non-audio tokens. We attribute this phenomenon to insufficient pretraining due to computational resource constraints, leading to partial task confusion between ASR and TTS during instruction fine-tuning. After filtering out non-audio tokens from AB-3’s generated sequences, we computed the CER metric and found that AB-3 still underperformed slightly compared to AB-1.

\section{Conclusion}
We present DistilCodec, an innovative audio codec distillation framework that consolidates multi-codebook architectures (RVQ/GVQ) into a singular high-capacity codebook (32,768 codes) while achieving near-perfect code utilization and mitigating mode collapse. Leveraging DistilCodec as our audio encoder, we develop UniTTS – a Qwen2.5-7B-based model trained through a tri-task pretraining regimen encompassing speech autoregression, text autoregression, and cross-modal speech-text generation, empirically demonstrating reduced dependence on precisely aligned text-speech data. Our experiments reveal UniTTS's capability to generate semantically coherent, emotionally expressive speech, with quantitative evidence showing that text instruction data enhances audio quality. Preliminary observations of text-audio modality competition suggest future exploration of MoE architectures and inter-modal optimization strategies.

\bibliographystyle{plain}
\bibliography{ref}


\appendix

\section{DistilCodec}
\subsection{Model Structure of DistilCodec} \label{app:distilcodec_structure}

\begin{figure}[htbp]
\centering
\includegraphics[scale=0.4]{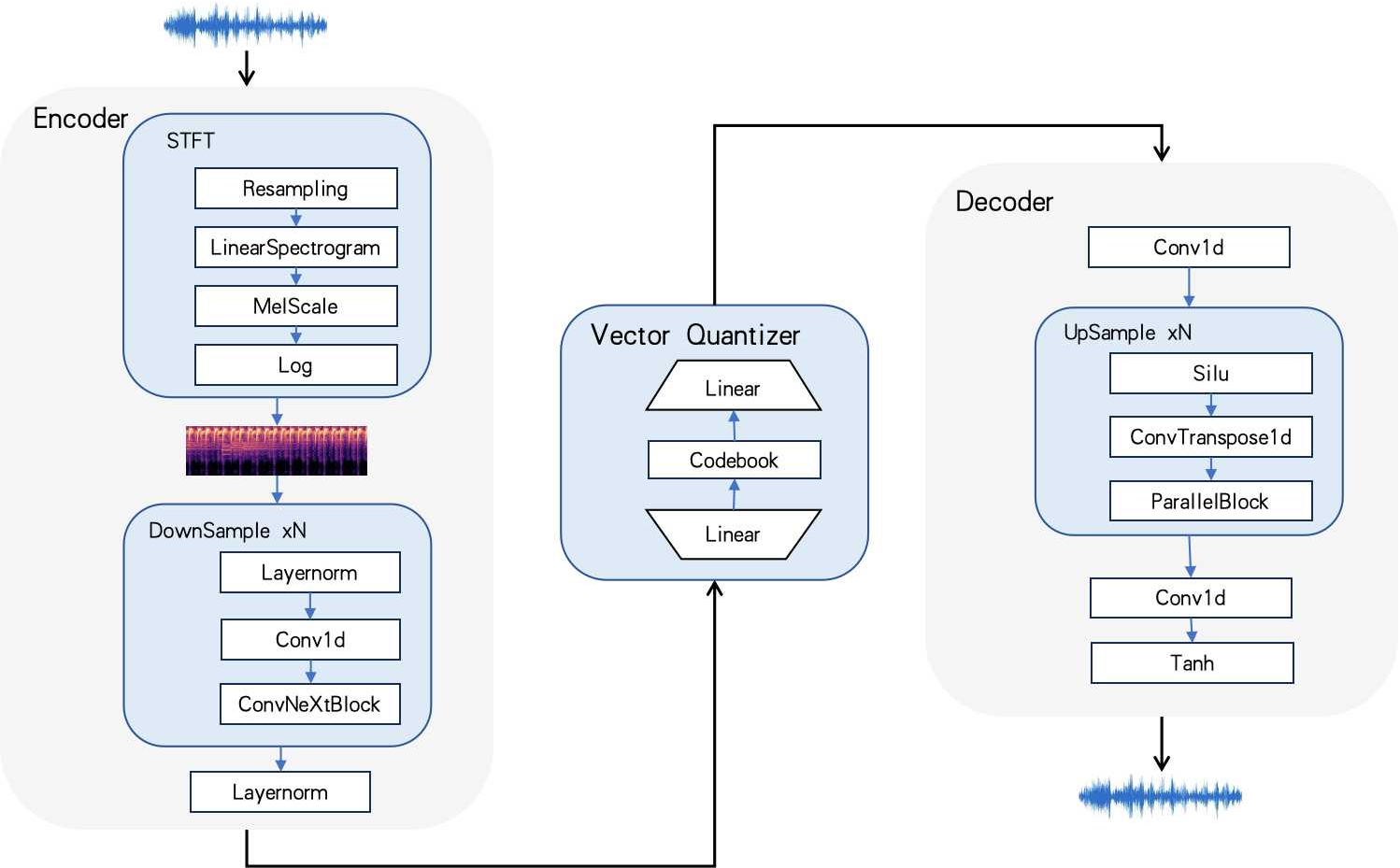}
\caption{The detailed network architecture of DistilCodec.}
\label{figure_4}
\end{figure}

The detailed network architecture of DistilCodec is illustrated in Figure \ref{figure_4}. DistilCodec consists of three key components:

\textbf{Audio Encoder}: The Mel-spectrogram parameters are specified in Table \ref{tab:stft_settings}, while the encoder configurations are detailed in Table \ref{tab:distilcodec_encoder_settings}.

\textbf{Vector Quantizer}: The quantizer and factor settings are provided in Table \ref{tab:vector_quantizer_settings}.

\textbf{Audio Decoder}: The specific parameters are listed in Table \ref{tab:decoder_settings}.

\begin{table}[h]
\centering
\caption{Mel-spectrogram STFT settings.}
\label{tab:stft_settings}
\begin{tabular}{l|c}
\toprule
\textbf{Configuration Item} & \textbf{Value} \\
\midrule
Sampling Rate & 24000 \\
Segment Size & 72000 \\
Mel Channels & 128 \\
Hop Size & 256 \\
Window Size & 1024 \\
FMin & 0 \\
FMax & 12000 \\
\bottomrule
\end{tabular}
\end{table}

\begin{table}[h]
\centering
\caption{Encoder settings of DistilCodec.}
\label{tab:distilcodec_encoder_settings}
\begin{tabular}{l|c}
\toprule
\textbf{Configuration Item} & \textbf{Value} \\
\midrule
Input Channels & 128 \\
Number of Downsampler & 4 \\
Downsampler Depth & [3, 3, 9, 3] \\
Downsampler Output Dims & [256, 512, 768, 1024] \\
ConvNeXt Drop Rate & 0.2 \\
Conv Kernel Size & 7 \\
\bottomrule
\end{tabular}
\end{table}

\begin{table}[h]
\centering
\caption{Settings for DistilCodec's Vector Quantizer.}
\begin{tabular}{l|c}
\toprule
\textbf{Configuration Item} & \textbf{Value} \\
\midrule
Number of Residual & 1 \\
Number of Group & 1 \\
Number of Codebook & 1 \\
Dimension of Code & 3584 \\
Number of Codes & 32768 \\
EMA Decay Rate & 0.8 \\
Conv Kernel Size & 7 \\
Pre Factorized Dense & [1024, 3584] \\
Post Factorized Dense & [3584, 1024] \\
\bottomrule
\end{tabular}
\label{tab:vector_quantizer_settings}
\end{table}

\begin{table}[h]
\centering
\caption{Decoder settings of DistilCodec.}
\begin{tabular}{l|c}
\toprule
\textbf{Configuration Item} & \textbf{Value} \\
\midrule
Number of Upsampler & 5 \\
Upsampler rates & [8, 4, 2, 2, 2] \\
Resblock Kernel Sizes & [3, 7, 11] \\
Resblock Dilation Sizes & [[1, 3, 5], [1, 3, 5], [1, 3, 5]] \\
ConvNeXt Drop Rate & 0.2 \\
Conv Kernel Size & 7 \\
Pre Conv1d Kernel Size & 13 \\
Post Conv1d Kernel Size & 13 \\
\bottomrule
\end{tabular}
\label{tab:decoder_settings}
\end{table}

\subsection{Discriminators of DisilCodec} \label{app:discriminators}
During the training phase, the GAN-based framework employs three discriminators:

\textbf{Multi-period discriminator}: Detailed parameters are provided in Table \ref{tab:multi_period_discriminator_settings}.

\textbf{Multi-scale discriminator}: Detailed parameters are listed in Table \ref{tab:multi_scale_discriminator_settings}.

\textbf{Multi-STFT discriminator}: Specific configurations can be found in Table \ref{tab:multi_stft_discriminator_settings}.

\begin{table}[h]
\centering
\caption{Multi Period Discriminator parameter settings of DistilCodec.}
\begin{tabular}{l|c}
\toprule
\textbf{Configuration Item} & \textbf{Value} \\
\midrule
Number of Period Discriminator & 5 \\
Periods & [5, 8, 13, 19, 30] \\
Kernel Size & 5 \\
Stride & 3 \\
\bottomrule
\end{tabular}
\label{tab:multi_period_discriminator_settings}
\end{table}

\begin{table}[h]
\centering
\caption{Multi Scale Discriminator parameter settings of DistilCodec.}
\begin{tabular}{l|c}
\hline
\textbf{Configuration Item} & \textbf{Value} \\
\midrule
Number of Period Discriminator & 5 \\
Periods & [5, 8, 13, 19, 30] \\
Kernel Size & 5 \\
Stride & 3 \\
\bottomrule
\end{tabular}
\label{tab:multi_scale_discriminator_settings}
\end{table}

\begin{table}[h]
\centering
\caption{Multi STFT Discriminator parameter settings of DistilCodec.}
\begin{tabular}{l|c}
\toprule
\textbf{Configuration Item} & \textbf{Value} \\
\midrule
Number of STFT Discriminator & 5 \\
Number of FFTs & [1024, 2048, 512, 256, 128] \\
Hop Lengths & [256, 512, 128, 64, 32] \\
Window Lengths & [1024, 2048, 512, 256, 128] \\
Number of Filter & 32 \\
\bottomrule
\end{tabular}
\label{tab:multi_stft_discriminator_settings}
\end{table}

\subsection{DistilCodec Training Framework} \label{app:DTF}
\begin{figure}[htbp]
\centering
\includegraphics[scale=0.4]{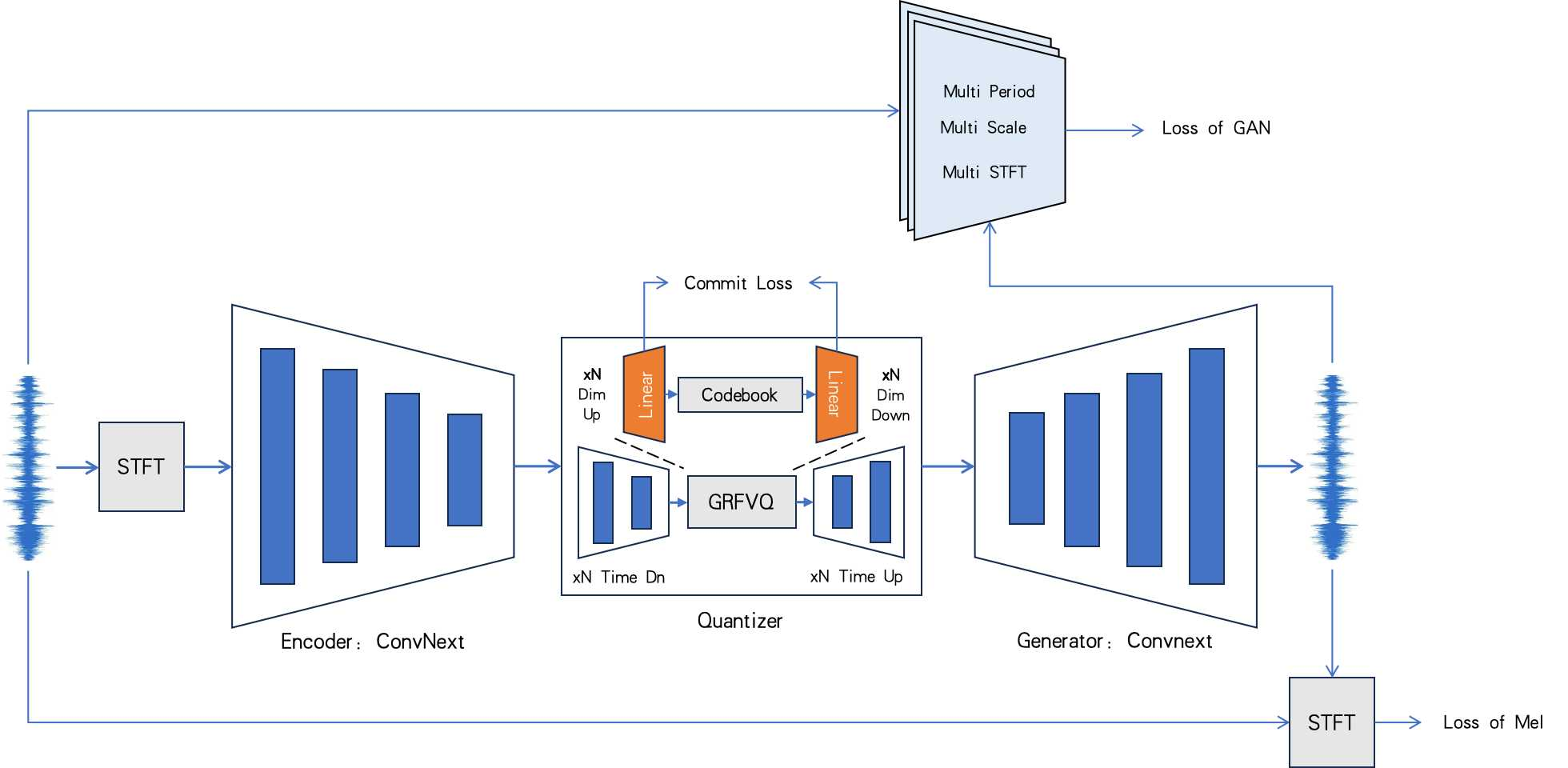}
\caption{Training Diagram of DisitilCodec.}
\label{figure_3}
\end{figure}
Fig.\ref{figure_3} illustrates the comprehensive training diagram of DisitilCodec.
The configuration of optimizer parameters during the training of DistilCodec can be found in Table \ref{tab:adamw_parameters}, while the LSGAN training pseudocode for DistilCodec (DLT) is outlined in Algorithm \ref{alg:lsgan_pseudo_code}. The training data of DistilCodec is illustrated in Table \ref{tab:codec_data_size}.
\begin{table}[h]
\caption{AdamW Experiment Parameter Settings}
\label{tab:adamw_parameters}
\centering
\begin{tabular}{l|c}
\toprule
\textbf{Parameter} & \textbf{Value} \\
\midrule
$\beta_1$ & 0.5 \\
$\beta_2$ & 0.9 \\
LR Decay & 0.98 \\
Weight Decay & 0.001 \\
\bottomrule
\end{tabular}
\end{table}

\begin{algorithm}
\centering
\caption{DistilCodec's LSGAN Training Pseudo Code}
\label{alg:lsgan_pseudo_code}
\begin{algorithmic}[l]
\Require $\text{Audio}_\text{universal}$
\For{epoch $= 0, 1, 2, \ldots$}
    \For{step $= 0, 1, 2, \ldots$}
        \State $e_f = \text{encoder}(x)$
        \State $(\text{quantized\_f}, l_{\text{commit}}) = \text{quantizer}(e_f)$
        \State $y_g = \text{generator}(\text{quantized\_f})$
        \State $l_{\text{mel}} = \text{multi\_scale\_mel\_loss}(y, y_g.\text{detach}())$
        \State $l_{\text{gan}} = \text{gan\_loss}(y, y_g)$
        \State $l_{\text{total}} = l_{\text{mel}} + l_{\text{commitment}} + l_{\text{gan}}$
        \State \Call{backward}{$l_{\text{total}}$}
    \EndFor
\EndFor
\Ensure Trained Audio Codec
\end{algorithmic}
\end{algorithm}

\begin{table}[h]
\caption{Distribution of DistilCodec training data}
\label{tab:codec_data_size}
\centering
\begin{tabular}{l|c}
\toprule
\textbf{Data Category} & \textbf{Data Size (in hours)} \\
\midrule
Chinese Audiobook & 38000 \\
Chinese Common Audio & 20000 \\
English Audiobook & 10000 \\
English Speech & 30000 \\
Music & 2000 \\
Total & 100000 \\
\bottomrule
\end{tabular}
\end{table}

\section{UniTTS} \label{app:unitts}
\begin{figure}[htbp]
\centering
\includegraphics[scale=0.5]{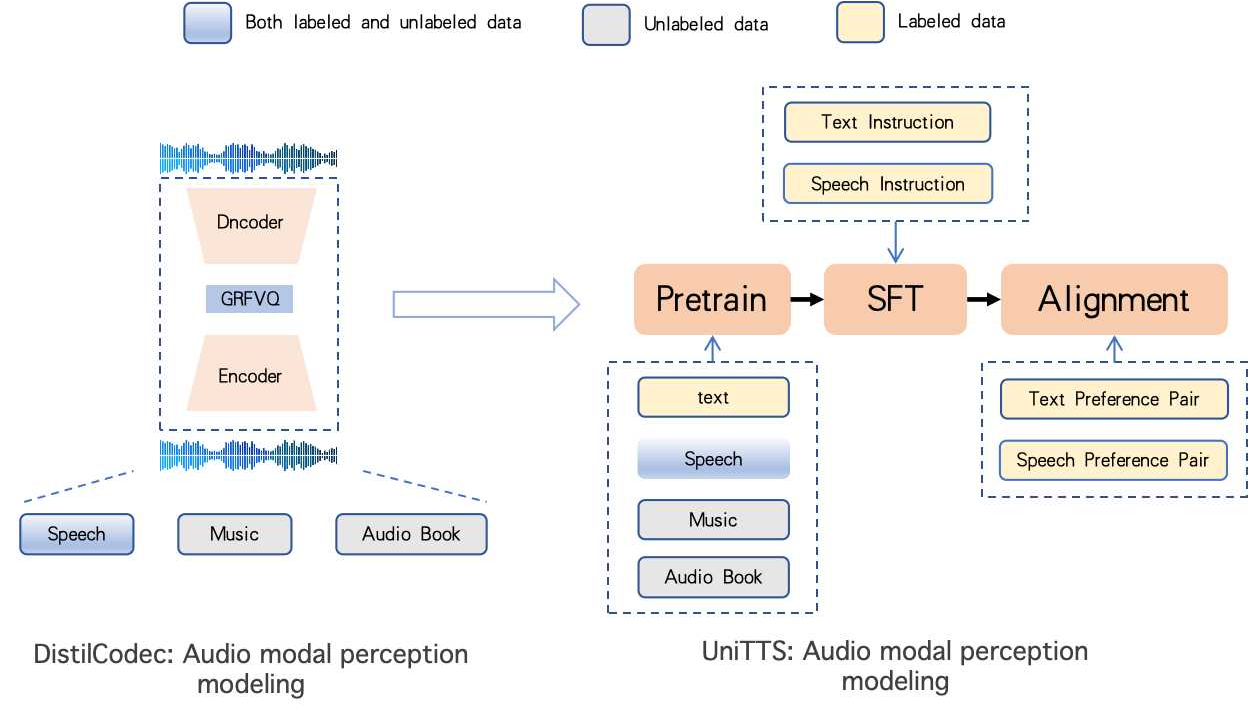}
\caption{Training schema of UniTTS and DistilCodec. DistilCodec consists of three core components: an Encoder, GRFVQ, and a Decoder, trained on universal audio data. The training process of UniTTS follows a methodology analogous to that of Large Language Models (LLMs), comprising three stages: Pretraining, Supervised Fine-Tuning (SFT), and Alignment. Notably, the pretraining phase utilizes universal audio as part of its training data.}
\label{system_training}
\end{figure}
Fig.\ref{system_training} illustrates the training schema of UniTTS and DistilCodec.

\subsection{UniTTS Prompt} \label{app:prompt_template}
\begin{figure}
\centering
\includegraphics[scale=0.35]{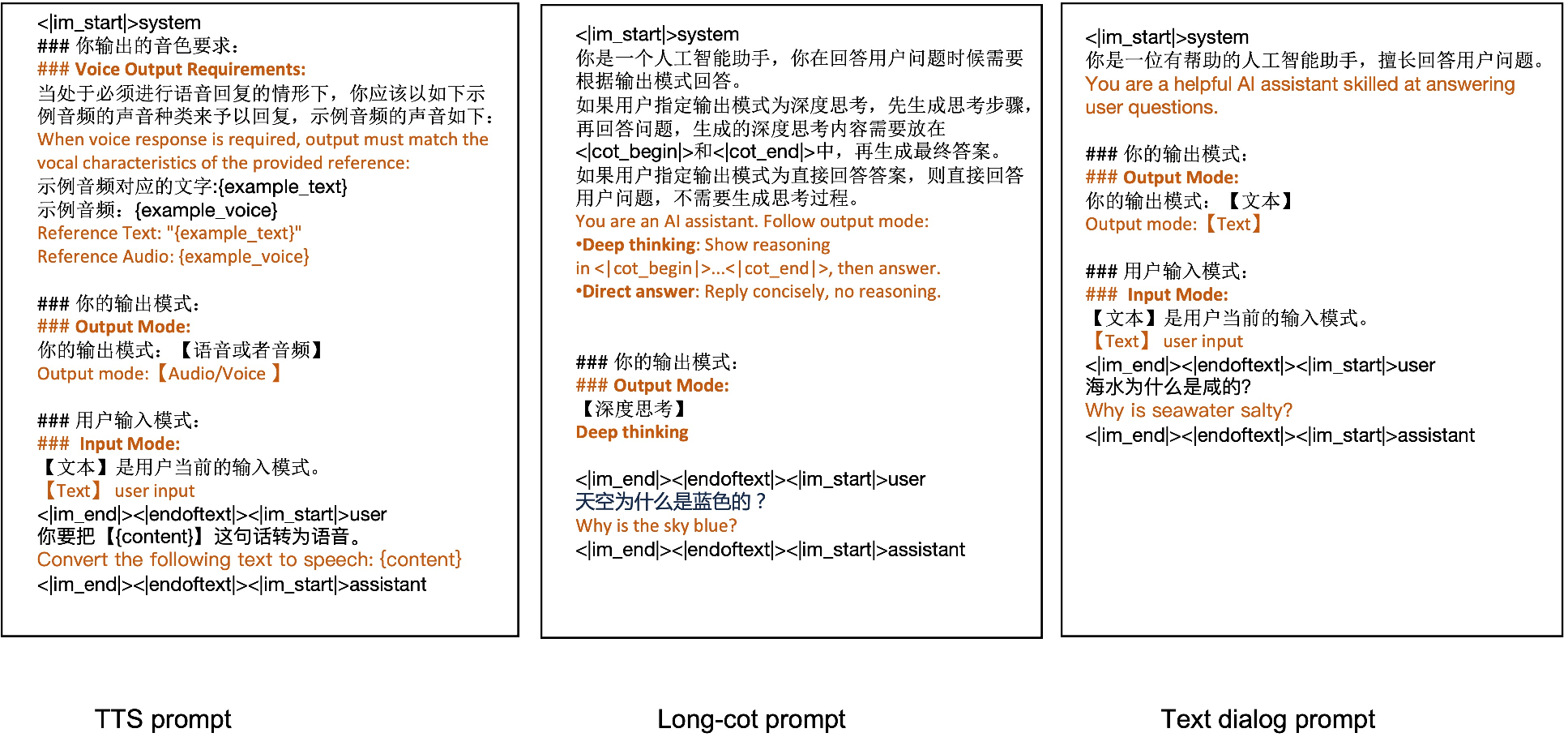}
\caption{Inference Prompt Template}
\label{figure_7}
\end{figure}

The TTS prompt template, long-CoT prompt template, and text dialogue template are detailed in Fig.\ref{figure_7}

\subsection{UniTTS Data Filtering Algorithm} \label{app:data_filtering}
The sft data filtering algorithm of UniTTS is illustrated in Algorithm \ref{alg:quality_filtering}.

\begin{algorithm}
\centering
\caption{The data filtering algorithm based on quality scores}
\label{alg:quality_filtering}
\begin{algorithmic}[1]
\Require Text-Audio Alignment Dataset $D(x, y)$ with $N$ samples
\Ensure scores
\State Initialize empty list \textit{scores} = []
\For{$i \gets 1$ to $N$}
    \State $x_i, y_i = D(i)$
    \State $dnsmos(i) = DNSMOS~P.835~OVRL(x_i, y_i)$
    \State $transcribed\_text1 = paraformer(y_i)$
    \State $transcribed\_text2 = whisper(y_i)$
    \State $cer(i) = cal\_cer(transcribed\_text1, transcribed\_text2)$
    \State $quality(i) = dnsmos(i) - cer(i)$
    \State $vad\_proportion = cal\_vad(y_i)$
    \If{$vad\_proportion > 0.14$}
        \State \textbf{continue}
    \EndIf
    \State scores.append($(x_i, y_i, quality(i))$)
\EndFor
\State Sort scores by $q$ in descending order $\triangleright$ In-place sort (high to low)
\State \Return scores
\end{algorithmic}
\end{algorithm}

\subsection{Distribution of Pretraining Data} \label{section:pertaining_data}

Distribution of pretraining data is illustrated in Table \ref{tab:pretrain_data_size}
\begin{table}[h]
\caption{Distribution of pretraining data}
\label{tab:pretrain_data_size}
\centering
\begin{tabular}{l|c}
\toprule
\textbf{Data Type} & \textbf{Data Size (B)} \\
\midrule
Text Data & 140 \\
Text-Audio Alignment Data & 82 \\
Audio Data & 100 \\
Total & 322 \\
\bottomrule
\end{tabular}
\end{table}

\subsection{Distribution of Sft Data} \label{section:sft_data_distrition}

\begin{table}[h]
\caption{Distribution of sft data}
\label{tab:sft_data_samples}
\centering
\begin{tabular}{l|c}
\toprule
\textbf{Data Type} & \textbf{Number of Samples} \\
\midrule
Text Data & 181K \\
Long-cot Dataset & 55K \\
Text-Audio Alignment Data & 401K \\
Total & 637K \\
\bottomrule
\end{tabular}
\end{table}

Distribution of sft data is illustrated in Table \ref{tab:data_sample_dpo_distribution}

\begin{table}[h]
\caption{Distribution of lpo data}
\label{tab:data_sample_dpo_distribution}
\centering
\begin{tabular}{l|c}
\toprule
\textbf{Data Type} & \textbf{Number of Samples} \\
\midrule
General SFT Data & 100K \\
Long-cot Dataset & 45K \\
Text-Audio Alignment Data & 300K \\
Total & 445K \\
\bottomrule
\end{tabular}
\end{table}

\begin{table}[h]
\caption{LPO Parameter values}
\label{tab:dpo_param_values}
\centering
\begin{tabular}{l|c}
\toprule
\textbf{item} & \textbf{value} \\
\midrule
$\beta$ & 0.2 \\
$\delta$ & 10.0 \\
$r_1$ & 1.0 \\
$r_2$ & 0.4 \\
Max LR & 8e-7 \\
Min LR & 5e-7 \\
Global Batchsize & 120 \\
\bottomrule
\end{tabular}
\end{table}

\subsection{Linear Preference Optimization} \label{app:lpo}
In formula \ref{lpo_eq}, $x_1^{\text{ste}}$, $x_2^{\text{ste}}$, $\gamma$, $x_1$, $x_2$ are shown in the following:
\begin{equation}
x_1^{\text{ste}} = r_1 \cdot \max(0, x_1 - x_2.\text{detach}() - \frac{1}{2\beta})
\end{equation}
\begin{equation}
x_2^{\text{ste}} = r_2 \cdot \max \left( 0, x_{1\text{.detach}()} - x_2 - \frac{1}{2\beta} \right)
\end{equation}
\begin{equation}
\gamma = 2\beta \cdot \frac{2}{r_1 + r_2}
\end{equation}
\begin{equation}
x_1 = \frac{\Pi_{\theta}(y_w|x)}{\Pi_{\text{ref}}(y_w|x)}
\end{equation}
\begin{equation}
x_2 = \frac{\Pi_{\theta}(y_l|x)}{\Pi_{\text{ref}}(y_l|x)}
\end{equation}
Here, $\Pi_{\theta}$ denotes the policy to be optimized, and $\Pi_{\text{ref}}$ represents the reference model, which is equivalent to the UniTTS model after Supervised Fine-Tuning (SFT). The variables $y_w$ and $y_l$ correspond to a pair of samples where, given the input prompt $x$, $y_w$ demonstrates superior performance compared to $y_l$.
LPO's hyperparameter is illustrated in Table \ref{tab:dpo_param_values}.

\subsection{Evaluation Criteria} \label{app:tts_evaluation_criteria}
To assess model performance, we conducted a Mean Opinion Score (MOS) evaluation on the test set constructed in Section 3.4 using the following criteria (rated on a 0-5 scale).

1) Fidelity: The audio accurately reproduces the original sound characteristics, including timbre and pitch alignment with ground truth recordings.

2) Stability: The audio playback exhibits no artifacts such as stuttering, frame skipping, or abrupt termination.

3) Naturalness: The output demonstrates human-like speech/instrument production without robotic artifacts or unnatural prosody.

4) Emotional expressiveness: The audio effectively conveys intended emotional states (e.g., joy, sadness, anger) with appropriate vocal/instrumental cues.

\subsection{Configuration of TTS Ablation Experiments}
\begin{figure}
\centering
\includegraphics[scale=0.4]{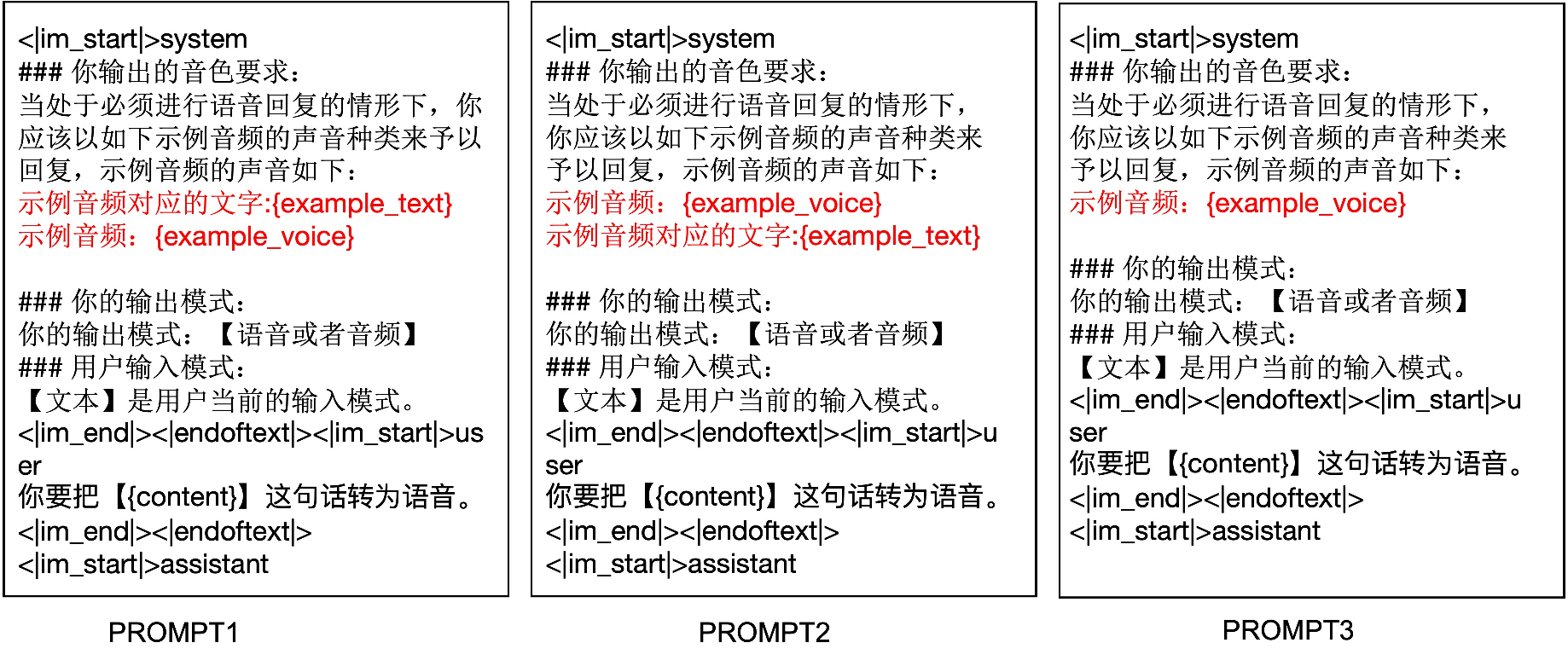}
\caption{Prompt configurations}
\label{figure5}
\end{figure}
\begin{table}[h]
\caption{Experimental settings for tts ablation tests}
\label{tab:tts_ablation_settings}
\centering
\begin{tabular}{c|l}
\toprule
\textbf{Ablation TestID} & \textbf{Experimental Setting} \\
\midrule
AB-1 & TTS dataset 1.2M, text dataset 181K, long-cot dataset 55K; prompt uses PROMPT1 \\
AB-2 & TTS dataset 1.2M, text dataset 181K, long-cot dataset 55K; prompt uses PROMPT3 \\
AB-3 & TTS dataset 1.2M, text dataset 181K, long-cot dataset 55K; prompt uses PROMPT2 \\
AB-4 & TTS dataset 1.2M; prompt uses PROMPT1 \\
AB-5 & TTS dataset 1.2M, text dataset 18.1w, long-cot dataset 55K; prompt uses PROMPT2 \\
\bottomrule
\end{tabular}
\end{table}
We define three prompt configurations: PROMPT1 presents both the exemplar audio and its corresponding text, with text displayed before audio; PROMPT2 reverses this order by placing audio before text; PROMPT3 omits the textual content while retaining the audio example from PROMPT1.

\subsection{Pre-training Loss of Stage 1} \label{Pre_training_Loss_of_Stage_1}
\begin{figure}
\centering
\includegraphics[scale=0.5]{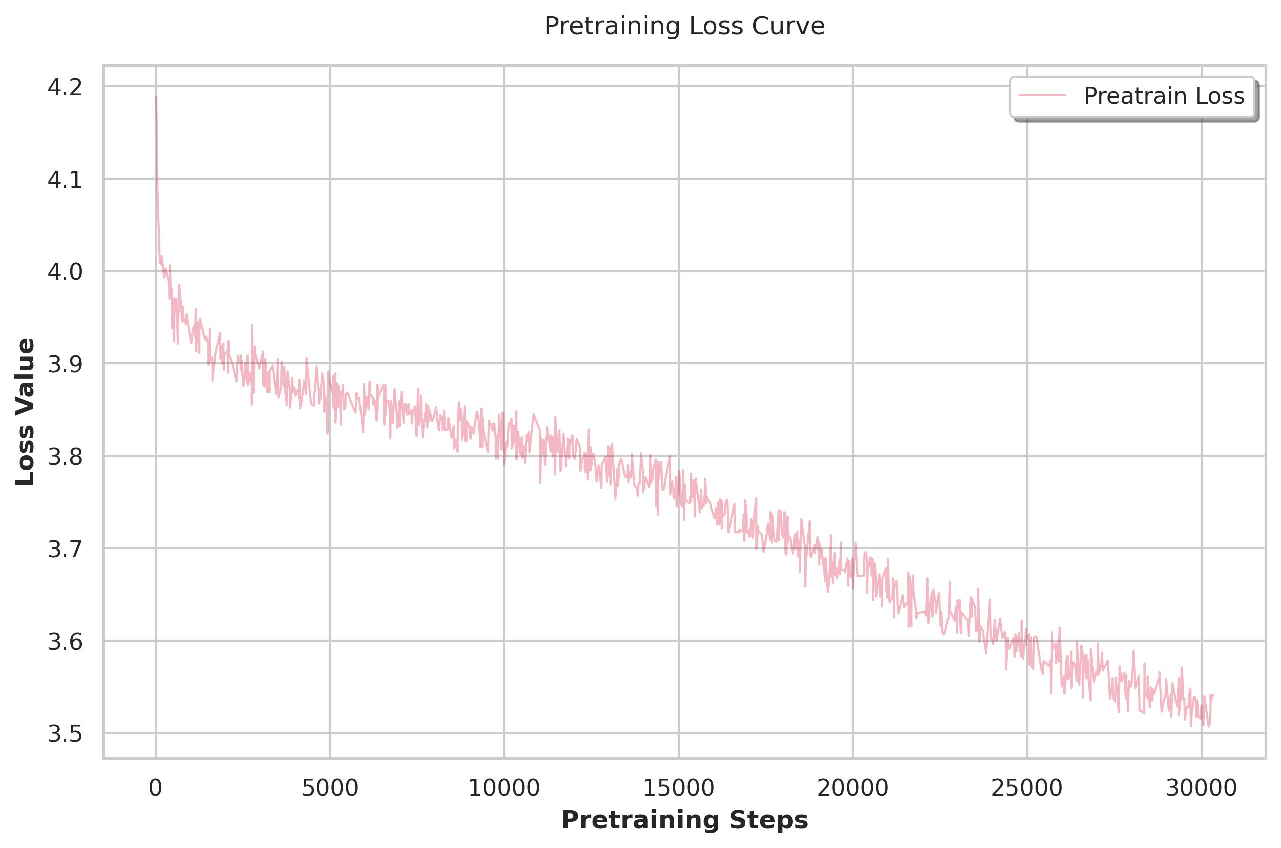}
\caption{Stage 1 Pre-training Loss}
\label{figure_8_training_loss}
\end{figure}
We present the loss curve from the first phase of pre-training. As shown in Fig.\ref{figure_8_training_loss}, the model's loss remains relatively high but exhibits a clear downward trend, indicating that the computational resources and training data during the pre-training stage were still insufficient.

\subsection{Text Capability Testing of UniTTS} \label{Text_Capability_Testing_of_UniTTS}
\begin{table}[h]
\caption{Performance comparison across different datasets and stages}
\label{tab:text_performance_comparison}
\centering
\begin{tabular}{l|c|c|c}
\toprule
\textbf{datasets} & \textbf{qwen2.5-7b-base-opencompass} & \textbf{pretrain-stage1} & \textbf{pretrain-stage2} \\ 
\midrule
MMLU & 74.26 & 47.12 & 52.44 \\
ARC-C & 59.66 & 32.0 & 37.97 \\
Winogrande & 68.98 & 59.12 & 58.96 \\
Hellaswag & 86.63 & 63.24 & 62.18 \\
GPQA & 39.39 & 23.23 & 25.76 \\
MATH & 51.2 & 2.86 & 7.78 \\
GSM8K & 79.45 & 19.18 & 64.97 \\
HumanEval & 77.44 & 10.98 & 14.63 \\
\bottomrule
\end{tabular}
\end{table}
The pretraining process consists of two stages. In Stage 1, we observed that incorporating the audio modality impaired text generation performance, resulting in overall task degradation due to model capacity constraints and modal competition. We hypothesize that enhancing the model's textual instruction-following capability could potentially improve both contextual comprehension and audio generation quality. This hypothesis was empirically validated in SFT e
xperiment \ref{ablation_test}, where improved audio generation outcomes were observed following the enhancement of text instruction capabilities. 

For Stage 2 pretraining, we implemented a strategic data augmentation approach by integrating text-based instruction datasets. This intervention successfully restored textual generation performance to a significant degree, as evidenced by the comparative analysis presented in Table \ref{tab:text_performance_comparison}. Nevertheless, it is important to note that code instruction and mathematics-related datasets were not included from the Stage 2 pretraining phase, which may account for the observed suboptimal performance in human evaluation metrics and mathematical reasoning tasks.

\subsection{Experimental Validation of the Text-to-Instruction Data Alignment Framework} \label{PA_enhance_audio}

\begin{table}[h!]
\caption{Comparison of CER for different TestIDs}
\label{tab:sft_cer_comparison}
\centering
\begin{tabular}{l|c}
\toprule
\textbf{TestID} & \textbf{CER} \\
\midrule
AB\_SFT & 18.18\% \\
UniTTS-SFT & 3.43\% \\
\bottomrule
\end{tabular}
\end{table}

Existing TTS models predominantly employ text-audio instruction fine-tuning approaches. In our study, we utilized 6.2 million text-audio pairs to evaluate the model's audio generation performance under full acoustic information modeling based on DistilCodec. The CER results of diffenrent TTS experiment setups are shown in Table \ref{tab:sft_cer_comparison}, which reveals a Character Error Rate (CER) of 18.18\% with suboptimal audio quality, and further analysis demonstrated that when processing identical text inputs, the model generated audio tokens with a mere 5\% repetition rate across two synthesis attempts. The result indicates that the model needs to handle a substantially larger audio sequence space (requiring comprehensive modeling of prosody, timbre, semantic information, etc.) compared to text sequence modeling, presenting significantly greater challenges.

Given the scarcity of high-quality text-audio alignment data, we drew inspiration from the training methodologies of unlabeled audio data in codec systems. During the cognitive modeling pre-training phase, we incorporated 100 billion unlabeled audio samples for training. Subsequently, in the Supervised Fine-Tuning (SFT) stage, using only 401k text-audio aligned samples, we achieved a significantly improved CER of 3.43\%.

\end{document}